\documentclass[twocolumn,amsfonts,showpacs,prl]{revtex4}
\usepackage{graphicx}
\usepackage{dcolumn}
\usepackage{amsmath}
\begin{document}
%\preprint{HEP/123-qed}
\title[Short Title]{New Class of Magnetoresistance Oscillations: Interaction of a Two-Dimensional Electron Gas with Leaky Interface Phonons}
\author{M. A. Zudov}
 \thanks{Present address: Stanford Picosecond Free Electron Laser Center, Stanford University, Stanford, CA 94305}
\author{I. V. Ponomarev}
\author{A. L. Efros}
\author{R. R. Du}
\affiliation{Department of Physics, University of Utah, Salt Lake City, UT 84112}
\author{J. A. Simmons}
\author{J. L. Reno}
\affiliation{Sandia National Laboratories, Albuquerque, NM 87185}
\author{(Received 2 May 2000)}
\affiliation{}
\begin{abstract}

We report on a new class of magnetoresis\-tance oscillations observed in a high-mobility two-dimensional electron gas (2DEG) in GaAs-Al$_x$Ga$_{1-x}$As heterostructures.
Appearing in a weak magnetic field ($B<$ 0.3 T) and only in a narrow temperature range (2 K $<T<$ 9 K), these oscillations are periodic in $1/B$ with a frequency proportional to the electron Fermi wave vector, $k_F$.
We interpret the effect as a magnetophonon resonance of the 2DEG with leaky interface-acoustic phonon modes carrying a wave vector $q=2k_F$.
Calculations show a few branches of such modes on the GaAs-Al$_x$Ga$_{1-x}$As interface, and their velocities are in quantitative agreement with the data.
\end{abstract}
\maketitle

There are several classes of transverse magnetoresistance (MR) oscillations known to exist in a two-dimensional homogeneous electron gas (2DEG).
The most common of these are the Shubnikov-de Haas oscillations (SdH), which arise from a magnetic field $B$-induced modulation of the density of states at the Fermi level $E_F$.
They become more pronounced with decreasing temperature $T$.
The magnetophonon resonance (MPR) \cite{mpr,tsui} is a source of another class of oscillations resulting from the absorption of bulk longitudinal optical phonons.
These resonances appear under the condition $\omega_{LO}=l\omega_c$, where $\omega_{LO}$ and $\omega_c=eB/mc$ are the optical phonon and cyclotron frequencies respectively, $l$ is an integer, and $m$ is the effective mass of the carriers.
These oscillations are only seen at relatively high $T\sim 100-180$K \cite{tsui}.
Both SdH and MPR are periodic in $1/B$, but the SdH frequency (reciprocal period) scales with electron density as $n_e$, whereas MPR is $n_e$-independent.

In this Letter, we report on a new class of MR oscillations \cite{zudov} observed in a high-mobility 2DEG in GaAs-Al$_x$Ga$_{1-x}$As heterostructures.
Unrelated to either of the above origins, these novel oscillations are still periodic in $1/B$, but they appear {\em only} in a narrow temperature range (2 K $<T<$ 9 K), and their frequency scales with $\sqrt {n_e}$.
We interpret the data in terms of a magnetophonon resonance mediated by thermally excited {\em leaky interface-acoustic phonon} (LIP) modes.
In principle, the surface modes might provide a good explanation as well, but in our case 2DEG is located so far from the surface ($\sim 0.5$ $\mu$m) that no such interaction is possible.

The leaky interface modes have been studied for a few decades in connection with the Earth's crust \cite{maradudin}.
The term ``leaky'' shows that the waves propagate at a small angle with the interface, so that the energy radiates away from the boundary.
For some specific parameters these waves may not be leaky \cite{stoneley}, but for the interface under study all of them are leaky.
Despite the fact that LIP is commonly present in layered material systems \cite{zinin}, it has so far not been considered on the GaAs-Al$_x$Ga$_{1-x}$As interface.
Due to radiation of energy, the frequency and velocity of leaky waves are complex:
$u=\omega/q=u_R-iu_I$ with $u_I \ll u_R$.

The novel oscillations can be explained by a simple momentum selection rule which is derived later in the paper.
It states that at high Landau levels (LLs) the electrons interact predominantly with the interface phonons carrying a wave vector $q=2k_F$, where $k_F$ is the Fermi wave vector of the 2DEG at zero $B$ field.
The condition for resonant absorption or emission of an interface phonon is then given by
\begin{equation}
 2k_F u_R= l \omega _c,\,\,\,\, \,\,\,\,l=1,2,3,...\,.
\label{con}
\end{equation}

We claim that Eq.~(\ref{con}) determines the values of $B$ for the {\em maxima} in these new MR oscillations.
It shows that the oscillations are periodic in $1/B$ with a frequency $f=2k_Fumc/e$.
Evidently, the bulk phonons can not account for the resonance, since their frequency depends on $q_z$, while the selection rule includes lateral momentum only.

Our primary samples are lithographically defined Hall bars cleaved from modulation-doped GaAs-Al$_{0.3}$Ga$_{0.7}$As heterostructures of high-mobility $\mu \approx 3\times 10^6$ cm$^2$/Vs.
The wafers are grown by molecular-beam epitaxy on the (001) GaAs substrate.
At low $T$, the density of the 2DEG, $n_e$ (in units of $10^{11}$ cm$^{-2}$ throughout the text), can be tuned by a combination of illumination from light-emitting diode and the NiCr front gate potential.
The experiments were performed in a variable-temperature $^4$He cryostat equipped with a superconducting magnet, employing a standard low-frequency lock-in technique for resistance measurement.

\begin{figure}[t]
\resizebox{0.47\textwidth}{!}{
\includegraphics{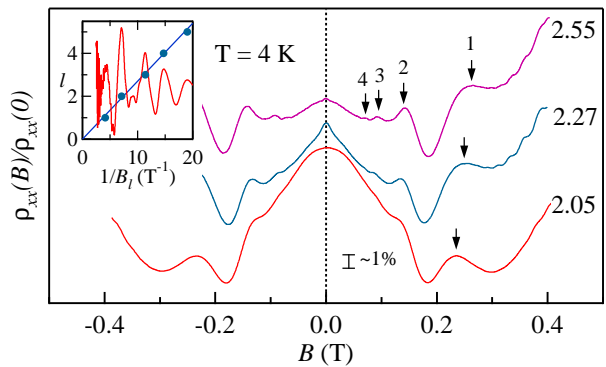}
}
%\centerline{\epsfxsize=3.3in\epsfbox{fig1c.eps}}
\caption{$\rho_{xx}(B)/\rho_{xx}(0)$ traces (shifted vertically for clarity) are shown for three densities $n_e$ of 2.05, 2.27 and $2.55\times 10^{11}$ cm$^{-2}$, respectively; arrows indicate the maxima for $l=1,\,2,\,3,\,4$ and the shift of the primary ($l=1$) peak with increasing $n_e$; Inset shows that the oscillations are periodic in $1/B$.}
\label{fig1}
\end{figure}

In Fig. \ref{fig1} we show the normalized low-field magnetoresistivity $\rho_{xx}(B)/\rho_{xx}(0)$ measured at $T = 4$ K, for the electron density $n_e=$ 2.05, 2.27, and 2.55, respectively.
In addition to the damped SdH commonly seen in a 2DEG at this $T$, the traces reveal new oscillations that appear only at $B<0.3$ T.
The amplitude of the oscillations is about 2-3 \% in these traces.
%%%%%%%%%%%%%%%%%%%%%%%%%%%%%%%%%%%%%%%%%%%%%%%%%
%%%%%%%%%%%%%%%%%%%%%%%%%%%%%%%%%%%%%%%%%%%%%%%%%%

Three aspects of the observation should be highlighted.
First, the oscillations are roughly periodic in inverse magnetic field, $1/B$.
The arrows next to the traces indicate the $\rho_{xx} (B_l)$ {\em maxima} (indexed as $l$ = 1, 2, 3, 4) in this oscillatory structure.
In the inset we plot the order of the oscillations, $l$ (and $-d^2\rho_{xx}/dB^2$), vs. $1/B$ for $n_e=$ 2.55 and observe a linear dependence.
Such periodic oscillations have been seen for all $n_e$ (from $\sim 1.5$ to 3) studied.
Second, with increasing $n_e$ the features shift orderly towards higher $B$.
Finally, the oscillatory structure is accompanied by a negative MR background, apparently in the same $B$ range where the oscillations take place.

We have now measured over a dozen specimen (from five wafers), of both the Hall bar (width from 10 $\mu$m to 500 $\mu$m) and the square (5 mm x 5 mm) geometries, and consistently observed similar oscillatory structures.
On the other hand, the significance of the ubiquitous MR background remains unclear.
Either negative or positive MR has been observed, and its strength (and even the sign) is largely specimen and cooling-cycle dependent.
In the following we shall focus on the analysis of the oscillatory structure, in particular, its dependence on $n_e$ and $T$.

\begin{figure}[t]
%\centerline{
%\epsfxsize=3.3in \epsfbox{fig2c.eps}
%}
\resizebox{0.47\textwidth}{!}{
\includegraphics{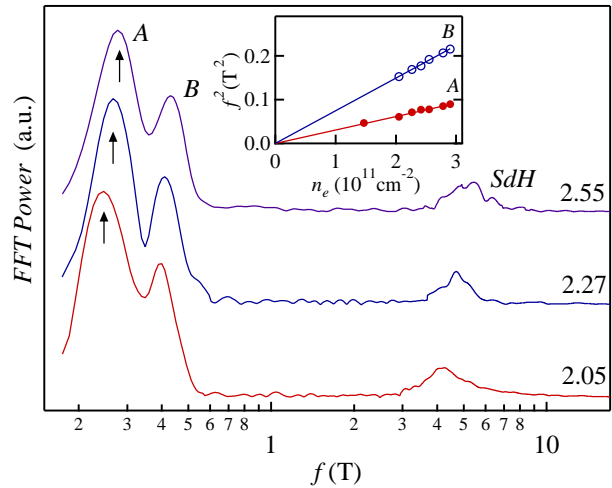}
}
\caption{Representative Fourier power spectra of the $\rho_{xx}(B)$ are shown for $n_e=$ 2.05, 2.27 and $2.55\times 10^{11}$ cm$^{-2}$, respectively, where the doublet (see text) is marked by $A$, $B$; arrows indicate the shift of the slow mode with increasing $n_e$.
Inset shows a linear dependence of $f_A^2$ (filled circles) and $f_B^2$ (open circles) vs. $n_e$; peak $A$ (peak $B$) positions are fitted with sound velocity $u_A=2.9$ km/s ($u_B=4.4$ km/s).}
\label{fig2}
\end{figure}

To further quantify our results, we have performed fast Fourier transform (FFT) on the resistance data.
As an example, Fig. \ref{fig2} shows the FFT power spectra obtained from the three traces in Fig. \ref{fig1} \cite{fft}.
Surprisingly, such analysis has uncovered two frequencies, marked by $A$ and $B$.
Peak $A$ corresponds to the main period, conforming to the simple fit in Fig. \ref{fig1}.
Peak $B$ is somewhat weaker, and occurs at $f_B\approx 1.5 f_A$.
The shift of the doublet with increasing $n_e$ is marked by three arrows for the main peak.
The FFT data have revealed a striking linear relation between the frequencies of oscillations and the electron Fermi wave vector.
We plot (see the inset) $f^2$ of the FFT peaks against the electron density, $n_e$, which has been varied from 1.47 to 2.95 in the same specimen.
Since $k_F =\sqrt{2\pi n_e}$, the observed linearity indicates that $f \propto k_F$.
%%%%%%%%%%%%%%%%%%%%%%%%%%%%%%%%%%%%%%%%%%%%%%
%%%%%%%%%%%%%%%%%%%%%%%%%%%%%%%%%%%%%%%%%%%%%%
Such a linear dependence distinguishes the new oscillations from SdH, as $f_{SdH} \propto k_F^2$, and is exactly what one expects from the phonon resonance scenario proposed here.
As such, the oscillatory structure must be viewed as resulting from the resonance of the 2DEG with two branches of the interface modes.
Using Eq. \ref{con} and a single known material parameter, the GaAs band electron mass $m \approx 0.068 m_e$, we fit the data (solid line in the inset) and deduce a velocity for the slow (fast) mode $u_A \approx$ 2.9 km/s ($u_B \approx 4.4$ km/s).
Within the experimental error of 10\% the data from several specimen collapse on the same lines, indicating that the new oscillations are generic in high-mobility 2DEG in GaAs-Al$_x$Ga$_{1-x}$As heterostructures.

\begin{figure}[t]
\resizebox{0.47\textwidth}{!}{
%\centerline{\epsfxsize=3.3in\epsfbox{fig3c.eps}}
\includegraphics{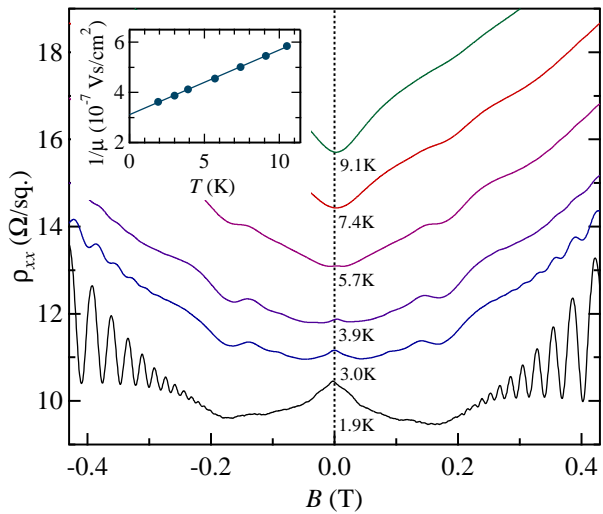}
}
\caption{$\rho_{xx}(B)$ (for $n_e = 2.27\times 10^{11}$ cm$^{-2}$) at
selected $T$ showing that oscillations are best developed at $T=3-7$ K
and are strongly suppressed at higher and lower $T$. Inset shows a
linear dependence of $1/\mu$ vs. $T$.}
\label{fig3}
\end{figure}

The $T$-dependence of the oscillations is consistent with a {\em thermally excited} phonon-scattering model.
Fig. \ref{fig3} shows the $\rho_{xx}(B)$ at selected temperatures (1.9 K $<T<$ 9.1 K ) where the evolution of the oscillations is clearly seen.
Notice first (see inset) that $\rho_{xx}(0)$ grows linearly with $T$, indicating that acoustic-phonon scattering dominates the electron mobility in this temperature range \cite{mendez,stormer}.
Considering the interface phonon modes of interest here, we use the value of the slow mode $u_A=2.9$ km/s to estimate a characteristic temperature, $T_c$, from $k_BT_c =\hbar u_A(2k_F)$.
The value of $T_c\approx$ 5 K can qualitatively account for the temperature dependence of the main features of the oscillations.
While the SdH gradually diminishes as $T$ increases, the oscillations are best developed at $T \approx 3-7$ K and are strongly damped at both higher and lower $T$.
At $T\ll T_c$ the number of interface phonons carrying $q=2k_F$ becomes small and therefore the amplitudes diminish.
At high $T$ the smearing of the LLs prevails and the oscillations disappear as well.
%%%%%%%%%%%%%%%%%%%%%%%%%%%%%%%%%%%%%%%%%%%%%%
%%%%%%%%%%%%%%%%%%%%%%%%%%%%%%%%%%%%%%%%%%%%%%

We now turn to the details of the theoretical explanation of the novel oscillations.
We have performed calculations \cite{efros} of LIP modes for the GaAs-Al$_{0.3}$Ga$_{0.7}$As interface on the basal (001) plane.
In the anisotropic case the speed of LIPs depends on angle between $q$ and the [100] direction.
Using the elastic moduli of the bulk lattices \cite{bulk} we found a series of modes with weak anisotropy and a small imaginary part of the velocity ($u_I/u_R<0.03$).
We have studied the modes within the interval of velocities 2.4$-$6.0 km/s.
Two close groups of modes have been found, one within the interval of 3$-$3.5 km/s and the other within 4.2$-$4.5 km/s.
These modes may be responsible for the two periods of oscillations which have been observed.
The frequencies of the other modes found are too high to be detected in our experiment.
Note that different modes may interact with electrons with different strengths.

To calculate the transverse conductivity due to the scattering of the 2DEG by the LIPs, we employ a 2D analog of the formula, first derived by Titeica \cite{titeica}:
\begin{eqnarray}\label{Titfor}
&&\sigma_{xx}=\frac{4\pi e^2}{A m^2 k_BT \omega _c^2}
\sum_{n,n'}\sum_{k_y,k_y'}\sum_{q_x,q_y}|I_{nn'}(q\lambda)|^2 q_y^2
|C(q)|^2\nonumber\\ &&\times
N_{l}f_{n}(1-f_{n'})\delta_{k_y-k_y'+q_y}\delta(\omega _c(n'-n)-qu).
\end{eqnarray}
Here $A$ is the area, $N_l=\left(\exp(\hbar\omega /k_BT)-1\right)^{-1}$, $f_{n}=\left(\exp\left((E_n-\mu)/k_BT\right)+1\right)^{-1}$, $\lambda=\sqrt{\hbar c/eB}$ is the magnetic length, and $|C(q)|^2\equiv v(q)/A$ is the square modulus of the 2DEG-LIP interaction, which has a power law dependence on $q$.
This formula can be interpreted in the following way.
A 2D electron in a magnetic field has a wave function which is a product of a plane wave in the $y$ direction and an oscillatory wave function, centered at the position $x_0=-c\hbar k_y/eB$: $\Psi=\exp(ik_y y)\phi_n(x-x_0)$, where $n$ is the LL index.
In the absence of scattering the electric current may flow only in the $y$-direction, providing the Hall effect.
A transverse conductivity appears because an electron transfers wave vector $q_y=k_{y}'-k_y$ to a scatterer.
This is equivalent to a jump in the $x$-direction at a distance $\Delta x_0=c\hbar q_y/eB$.
In Eq. (\ref{Titfor}) this physics is applied to electron-interface phonon scattering.
The mechanism of the 2DEG-LIP interaction, which may be either deformation potential or piezoelectric interaction, is not particularly important for our purpose.
The overall scattering is of the same order as the bulk phonon scattering, since the energy densities of both excitations are of the same order in the vicinity of the interface.

\begin{figure}[t]
%\centerline{\epsfxsize=3.3in\epsfbox{fig4c.eps}}
\resizebox{0.47\textwidth}{!}{
\includegraphics{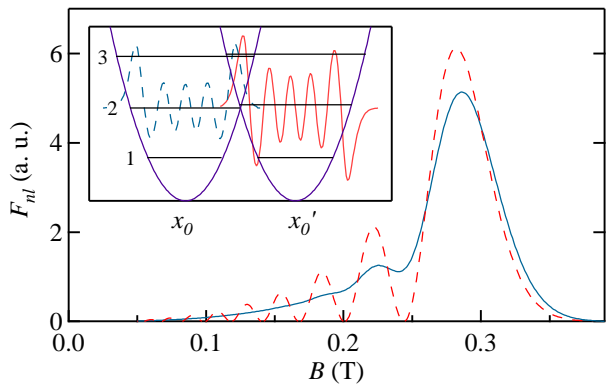}
}
\caption{ $F_{17,1}(B)$ in the limit $\sigma=0$ (dashed line) and for
LIP (solid line). Inset illustrates the origin of the strongest peak
in $F_{nl}(B)$.}
\label{fig4}
\end{figure}

If the interface phonon has no attenuation, the square of matrix element $I_{nn'}$ is given by \cite{sdh}
\begin{eqnarray}\label{matel}
|I_{n,n+l}(b)|^2 &=& \left|\int_{-\infty}^{+\infty}e^{iq_x x}\phi_n(x-x_0) \phi_{n+l}(x-x_0')\,dx\right|^2\nonumber\\
&=&\frac {n!}{(n+l)!} \left(\frac{b^2}{2}\right)^{l} e^{-\frac{b^2}{2}}
\left[L_n^l\left(\frac{b^2}{2}\right)\right]^2,
\end{eqnarray}
where $b=q\lambda$ and $L_n^l(x)$ is the generalized Laguerre polynomial.
Substituting summation over wave vectors by integration in Eq. (\ref{Titfor}) one obtains
\begin{eqnarray}
\label{interm}
&&\sigma_{xx}=\frac{e^2}{2\pi\hbar}\frac{1}{m k_BT \omega _c}
\sum_{n,l}N_{l}f_{n}(1-f_{n+l})\nonumber\\ && \times
\int_0^{\infty}dq\, q^3 v(q)\left| I_{n\,n+l}(q\lambda) \right|^2
\delta\left(\omega _c l-qu\right).
\end{eqnarray}
Taking into account the imaginary part of the LIP frequency, $\omega=q(u_R-iu_I)$, we can substitute for the $\delta$-function in Eq. (\ref{interm}) a Gaussian distribution with appropriate dispersion $\sigma=qu_I$.
Since the dispersion is small we can set $q=\omega _c l/u_R$ everywhere except for the strongly oscillating function $|I_{nl}(q\lambda)|^2$.
Then after averaging we obtain for the transverse conductivity
\begin{equation}
\label{fineq}
\sigma_{xx}=\frac{e^2/2\pi\hbar}{mu\omega _c k_BT}
\sum_{l,n}v\left(\frac {\omega _c l}{u}\right) F_{nl} N_l
f_n(1-f_{n+l}),
\end{equation}
where the function $F_{nl}$ can be expressed as a series of Hermite polynomials of imaginary argument:
\begin{eqnarray*}
F_{nl} & = & \frac{(\omega_c l/u)^3}{\sqrt{1+\alpha^{-1}}} \exp\left(-\frac{\alpha}{1+\alpha}\frac{\hbar\omega _c l^2}{2mu^2}\right) \nonumber \\
& &\times \sum_{k,j}^n \frac{n!(-1)^l}{(n+l)!k!j!}
 \left(\begin{matrix} {n+l} \\ {n-k} \end{matrix} \right)
 \left(\begin{matrix} {n+l} \\ {n-j} \end{matrix} \right)
% \left(\begin{array}{c} {n+l} \\ {n-j} \end{array} \right)
% \left( \begin{array} {c} {n+l} \\ {n-k} \end{array} \right)
 \nonumber \\
& &\times \frac{H_{2(k+j+l)}\left(il\sqrt{\frac{\hbar\omega_c}{2mu^2}}\frac{\alpha} {\sqrt{1+\alpha}}\right)}{[2\sqrt{1+\alpha}]^{2(k+l+j)}},
%\binom {n+l}{n-k} \binom {n+l}{n-j}
%\frac{H_{2(k+j+l)}\left(il\sqrt{\frac{\hbar\omega_c}{2mu^2}}\frac{\alpha} {\sqrt{1+\alpha}}\right)}{[2\sqrt{1+\alpha}]^{2(k+l+j)}},
\end{eqnarray*}
with $\alpha=(u/\sigma \lambda)^2$.
Hereafter, we assume that $u$ is the real part of the LIP velocity.
In Fig. \ref{fig4} we plot $F_{nl}$ for $n=17$ and $l=1$ as a function of $B$ for LIP with $\sigma =\omega_c u_I/u_R$ (solid line) and in the limit $\sigma =0$ (dashed line).
%%%%%%%%%%%%%%%%%%%%%%%%%%%%%%%%%%%%%%%%%%%%%%%%%%%%%
%%%%%%%%%%%%%%%%%%%%%%%%%%%%%%%%%%%%%%%%%%%%%%%%%%%
As we can see, once attenuation is introduced, only one strong peak remains that corresponds to Eq. (\ref{con}) at $l=1$.
This means that only phonons with wave vector $q=2\sqrt{2n}/\lambda$ effectively interact with electrons under the condition $n\gg 1$.
In fact, due to the Fermi distribution in Eq. (\ref{fineq}) only the values of $n\approx E_F/\hbar\omega_c$ are important.
Then, indeed, $2\sqrt{2n}/\lambda= 2\sqrt{2\pi n_e}=2k_F$, and we arrive at Eq. (\ref{con}) for $l=1$.
The same conclusion holds for any $l\ll n$.

This is an important result of our work. It can be interpreted from the following semi-classical consideration (see the inset in Fig. \ref{fig4}).
Let us consider $n\sim n' \gg 1$. Since the square of the matrix element in Eq. (\ref{matel}) depends on $q$ only, we can put $q_x=0$.
Then the integrand in (\ref{matel}) is an overlap of two oscillatory wave functions shifted with respect to each other.
In the vicinity of the turning point the wave function always has a maximum since the momentum is small and the particle spends most of its time there.
There are three possibilities.
Cases 1 and 3 in the inset show situations when turning points are apart from each other, and case 2 occurs when the turning points coincide in space.
Obviously, in case 2 the overlap integral has a maximum.
This occurs when $m\omega_c^2(\Delta x_0)^2/8=n\hbar\omega_c$, which is equivalent to the above condition $q\lambda\approx 2\sqrt{2n}$.
Note that the other maxima in Fig. \ref{fig4} can be smeared very easily because their widths are proportional to $n^{-1/2}$, while the first maximum near the turning point can be approximated by an Airy function and its width is independent of $n$.
Thus, the maxima in $F_{nl}(B)$ for different $l$ give rise to oscillations in $\rho_{xx}(B)$.

As a whole, the results provide good agreement with our experimental data.
In particular, the slow mode $u_A$ can be identified with the lower bunch of modes calculated here. Within the experimental uncertainty we are unable to find any anisotropy for the velocity, therefore, the data must be viewed as an average over all directions.
Likewise, the velocities of fast modes coincide with $u_B=$ 4.4 km/s, but this should be taken with caution.
Since our experiments have so far been centered on a temperature range around 5 K, a positive identification of the fast mode awaits for a more detailed $T$-dependence study at higher temperatures.

In conclusion, we have discovered a new class of mag\-ne\-to-oscil\-lations in a high-mobility 2DEG and interpreted it as a magnetophonon resonance with leaky interface-acoustic phonons.
Owning to their 2D characteristics, the leaky interface modes play a unique role in the scattering of 2D electrons in GaAs-Al$_x$Ga$_{1-x}$As heterostructures and quantum wells.
This role has never been studied before.

The experimental work (M.A.Z. and R.R.D.) is supported by NSF grant DMR-9705521. R.R.D. also acknowledges an Alfred P. Sloan Research Fellowship and thanks M. E. Raikh for helpful conversations.
The theoretical work (I.V.P. and A.L.E.) is supported by a seed grant of the University of Utah.
A.L.E. is grateful to R. L. Willett for insightful discussions.
The work at Sandia is supported by the US DOE under contract DE-AC04-94AL85000.


\begin{thebibliography}{}
\bibitem{fft}The FFT was performed on $-d^2\rho_{xx}/dB^2$ to enhance the signal/noise ratio. The broader SdH width reflects the fact that the data were from a limited $B$ range 0 $<B<$ 0.4 T.
\end{thebibliography}
\end{document}